\documentclass[manuscript]{aastex}
\def\msun{M_{\odot}}
\usepackage{graphicx}
\usepackage{natbib}
\usepackage[usenames]{color}
\begin{document}


\title{An AMR Study of the Common Envelope Phase of Binary Evolution}

\author{Paul M. Ricker\altaffilmark{1}
\email{pmricker@illinois.edu}
\and
Ronald E. Taam\altaffilmark{2,3}
\email{r-taam@northwestern.edu}}
\altaffiltext{1} {Department of Astronomy, University of Illinois, 1002 West Green Street,
Urbana, IL 61801}
\altaffiltext{2}{Northwestern University, Department of Physics and Astronomy,
2131 Tech Drive, Evanston, IL 60208}
\altaffiltext{3}{Academia Sinica Institute of Astrophysics and Astronomy-TIARA,
P.O. Box 23-141, Taipei, 10617 Taiwan}


\begin{abstract}
The hydrodynamic evolution of the common envelope phase of a low mass binary
composed of a $1.05 \msun$ red giant  and a $0.6 \msun$ companion has been
followed for five orbits of the system using a high resolution method in three
spatial dimensions.  During the rapid inspiral phase, the interaction of the
companion with the red giant's extended atmosphere causes about 25\% of the
common envelope to be ejected from the system, with mass continuing to be lost
at the end of the simulation at a rate $\sim 2 \msun {\rm\ yr}^{-1}$. In the process
the resulting loss of angular momentum and energy reduces the orbital separation by a 
factor of seven. After this inspiral phase the eccentricity of the orbit rapidly 
decreases with time.  The gravitational drag dominates hydrodynamic drag at all times 
in the evolution, and the commonly-used Bondi-Hoyle-Lyttleton prescription for estimating 
the accretion rate onto the companion significantly overestimates the true rate. On scales
comparable to the orbital separation, the gas flow in the orbital plane in the vicinity of 
the two cores is subsonic with the gas nearly corotating with the red giant core and 
circulating about the red giant companion. On larger scales, 90\% of the outflow is 
contained within 30 degrees of the orbital plane, and the spiral shocks in this material 
leave an imprint on the density and velocity structure. Of the energy released by the
inspiral of the cores, only about 25\% goes toward ejection of the envelope.
\end{abstract}


\section{Introduction}

\medskip

The common-envelope (CE) phase of binary star evolution, a brief phase during which two stars 
spiral together within a differentially rotating envelope leading to the loss of mass and orbital 
angular momentum from the system, was first proposed as an important stage in the evolution
of close binaries during the 1970s \citep{paczynski_common_1976}. Because close binary systems are 
central to the interpretation and understanding of many astronomical phenomena, including Type Ia 
supernovae, cataclysmic variables, X-ray binaries, and possibly short-duration gamma-ray bursts, 
developing a physical understanding of the CE phase has become important for predicting the formation 
rates and characteristics of these systems \citep{2000ARA&A..38..113T}. Mapping the portions of 
binary parameter space that lead to different CE outcomes is a particularly important goal. 

A definitive understanding of the CE phase requires computationally intensive hydrodynamical 
simulations of which a number of recent studies have been carried out 
\citep[e.g.,][]{ricker_interaction_2008,2010NewAR..54...65T,passy_ce}.
However, the final state of the post-CE systems has yet to 
be determined from such calculations and, hence, modeling of its outcomes in binary population 
synthesis studies has relied upon simple analytical prescriptions based on energetic considerations 
\citep{1985ApJS...58..661I} where an efficiency for the conversion of orbital energy of the binary to the 
kinetic energy of the outflow is assumed.  Alternatively, a prescription based on angular momentum 
considerations has been proposed \citep{2000A&A...360.1011N,2005MNRAS.356..753N}, where the specific 
angular momentum of the ejected matter is taken as a parameter.  Recent studies have focused on 
improvements to the former analytical prescription by developing a more accurate description of the 
conditions required for ejection: that is, the binding energy of the envelope and the response 
of the stellar interior to mass loss \citep{2010ApJ...716..114X,2010Ap&SS.329..243G,2010ApJ...719L..28D,2010arXiv1009.5400L}. 
However, the efficiency of the ejection process remains to be determined. 

Toward the goal of understanding the physical mechanisms that are important in governing the ejection 
of material during the CE phase and ultimately the efficiency of the process, we have carried out a 
high-resolution adaptive mesh refinement (AMR) simulation of the evolution of a binary initially 
consisting of a 1.05$\msun$ red giant containing a 0.36$\msun$ degenerate core with a 0.6$\msun$ 
companion. The use of AMR in this study represents a major improvement over previous work based on 
stationary nested grids \citep{1998ApJ...500..909S}.  This method ensures that the deep interior of the 
common envelope is always well resolved as matter moves about the center of mass of the system. Hence, 
the method permits one to study systems with binary components of nearly equal masses, which could not
be adequately modelled with the stationary nested grid technique.  In these systems, the CE phase is 
initiated by Roche lobe overflow of the red giant star rather than by a tidal (Darwinian) instability 
of the orbit.  The results from the first 41 days of our simulation were described in Ricker and Taam 
(\citeyear{ricker_interaction_2008}, hereafter Paper~I), wherein we showed that the assumption of 
Bondi-Hoyle-Lyttleton \citep[BHL; ][]{hoyle_effect_1939,bondi_mechanism_1944, bondi_spherically_1952} 
accretion onto the companion dramatically overestimates the true accretion rate since the conditions for 
BHL accretion (uniform supersonic motion with the gravitational wake trailing the motion of the accretor) are 
not satisfied.

In this paper we examine the intermediate time evolution of the simulation whose early stages were 
described in Paper~I. Here, we focus specific attention on the determination of the drag forces and 
the mechanisms for angular momentum transport in the system. To gain a deeper understanding of the 
influence of the interaction on the core of the red giant and the companion, we also study the 
character of the mass flow in the vicinity of each component and the accretion onto each component. 
In addition, we examine the angular and radial distribution of the matter ejected from the CE, which 
may provide observational evidence that a system has passed through this stage, independent of the 
outcome.  In \S 2 we briefly discuss the numerical methods and initial conditions used in our setup. 
The results are presented in \S 3, and we conclude with a discussion of their implications in the last 
section. 

\section{Numerical Methods}

We use a one-dimensional stellar structure and evolution code initially due to Eggleton 
(\citeyear{eggleton_evolution_1971,eggleton_composition_1972}) to create a model red giant star with 
total mass $1.05 \msun$, of which $0.69 \msun$ is contained in the extended gaseous envelope.  This 1D 
model specifies density, pressure, and composition as functions of radius. We linearly interpolate this 
model onto a 3D Eulerian mesh for use in the FLASH 2.4 adaptive mesh refinement (AMR) simulation code 
\citep{fryxell_flash:adaptive_2000}.  FLASH uses an oct-tree refinement scheme; we select blocks for 
refinement using the standard second-derivative criterion applied to gas pressure and temperature,
except in blocks for which the maximum gas density is less than $10^{-6}$~g~cm$^{-3}$.  The Euler 
equations of hydrodynamics are solved using the piecewise-parabolic method 
\citep[PPM;][]{colella_piecewise_1984} with modifications for nonideal equations of state based on 
\cite{colella_efficient_1985}.  The stellar equation of state that is used is based on tabulated values 
of the Helmholtz free energy and its derivatives for a three-isotope composition including hydrogen,
helium, and carbon \citep{timmes_accuracy_1999}.  The Poisson equation is solved using an adaptive 
multigrid method. Outflow boundaries are used for the hydrodynamics, while a variant of James' 
(\citeyear{james_solution_1977}) method is used to impose isolated boundary conditions on the 
gravitational field.
  
The 1D model is characterized by a radius of $2.2\times10^{12}$~cm, outside of which we augment it with 
a uniform ambient medium at the pressure of the outermost layer of the star and a density of 
$10^{-9}$~g~cm$^{-3}$ in order to avoid artificial outflow associated with inadequate resolution 
of the outermost layers. Within a box of size $4\times10^{13}$~cm, we apply enough levels of refinement (9, with $8^3$ zones per mesh block)
to achieve a smallest zone spacing of $2\times10^{10}$~cm. The innermost $0.36 \msun$ ($6\times10^{10}$ 
cm) of the gas in the red giant star is replaced by a spherical cloud of 200,000 particles which moves 
as a solid body and interacts with the gas only gravitationally. The resulting 3D model is allowed to
relax for one dynamical time (13 days). During this time, a damping factor is applied to the gas 
velocities at the end of each timestep, beginning with 0.9 and gradually reducing the amount of damping
by increasing the factor toward unity.  By the end of the relaxation stage the typical velocities 
are less than 2~km~s$^{-1}$. 
 
Once we have relaxed the 3D red giant model on the FLASH mesh, we introduce a $0.6 \msun$ companion, 
also represented by a spherical cloud of 200,000 particles moving as a solid body with radius 
$6\times10^{10}$~cm, in a circular orbit at an orbital separation of $4.3\times10^{12}$~cm.  The red 
giant is also set into orbital motion and is given a uniform spin, with spin angular velocity equal to 
95\% of the synchronous value. This corresponds to an initial surface velocity for the red giant of 
about 35~km~s$^{-1}$, which is within the observed range for red giants in short-period binaries 
\citep{massarotti_rotational_2008}.

The 3D CE calculation was carried out over a period of time on successive Teragrid platforms beginning with the Tungsten cluster at the National Center for Supercomputing Applications and ending with the Ranger cluster at the Texas Advanced Computing Center. Using the Teragrid service unit conversion calculator, we estimate the cost of the run up to 56.7 days of simulation time to be the equivalent of 512 cores for 792,000 core-hours on Ranger.

The description of the early evolution of the CE phase was briefly presented in Paper~I.  Here, we report on the further evolution of the system during which which envelope stripping accelerates and the orbit shrinks rapidly.  By 41~days the orbital separation has been reduced by a factor of 7, primarily because of the gravitational drag.  In the next section we discuss the intermediate time evolution beginning at this point.


\section{Results}


\subsection{Orbital evolution}

\begin{figure}
\centerline{\includegraphics*[height=3in]{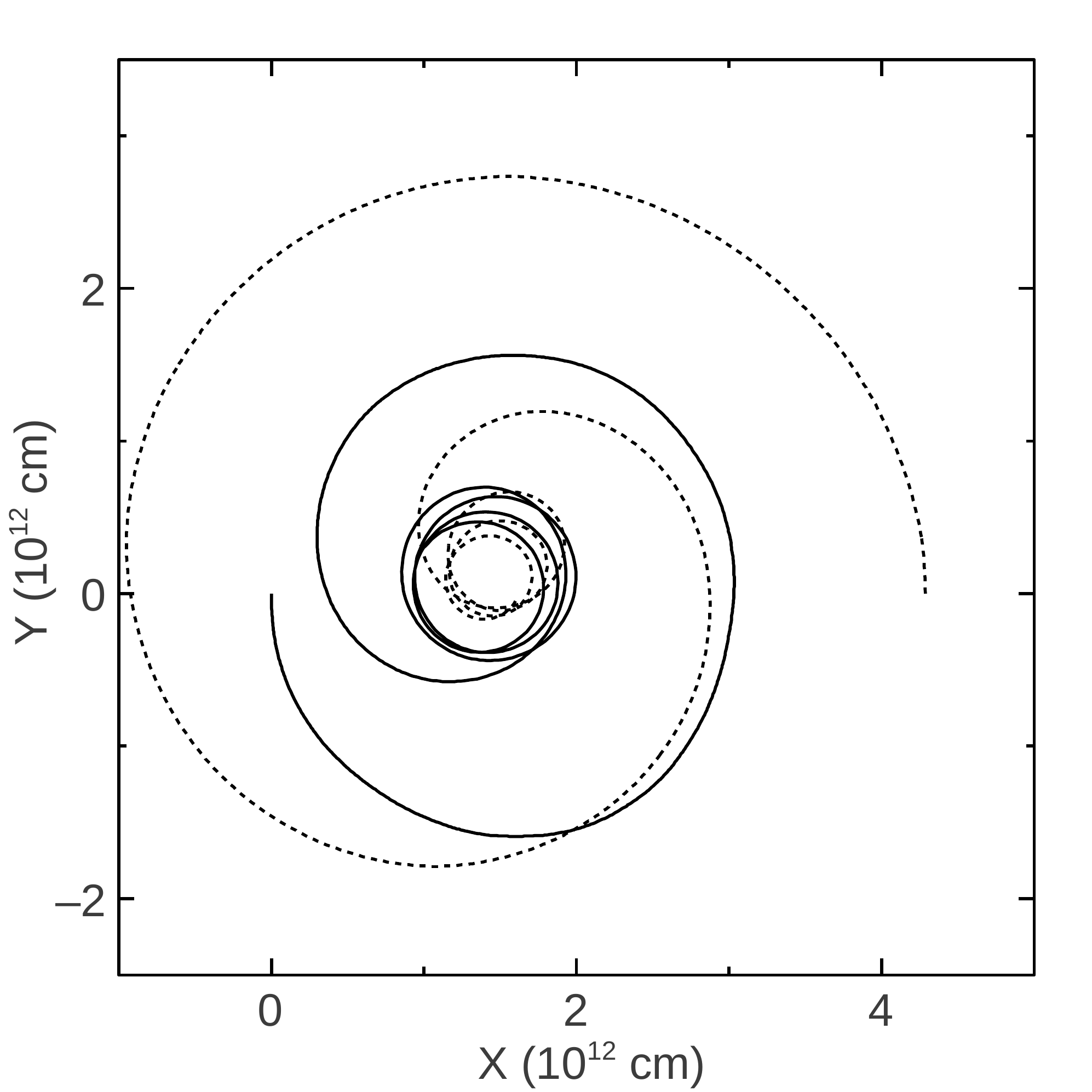}}
\caption{\label{Fig:orbit} Orbital trajectories of the red giant core (solid curve) and its companion 
(dashed curve) in the equatorial plane during the inspiral phase of the CE binary system.}
\end{figure}

\begin{figure}
\centerline{\includegraphics*[height=3in]{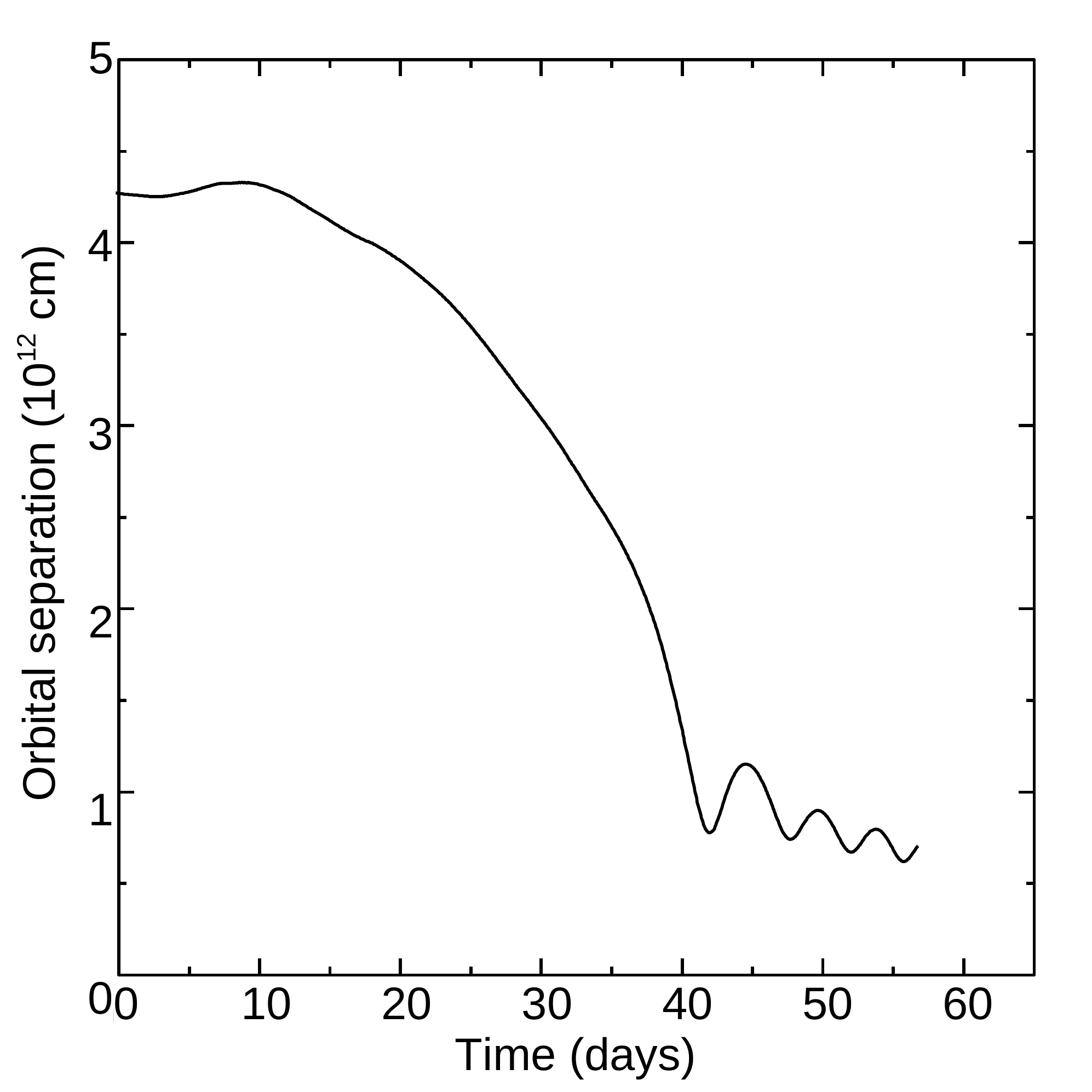}}
\caption{\label{Fig:separation} Orbital separation of the red giant and its companion as a function 
of time.}
\end{figure}

The evolution of the CE binary system described in the previous section has now been continued 
through 56.7 days of simulation time, or about five orbits.  Figures~\ref{Fig:orbit} and 
\ref{Fig:separation} illustrate the orbital trajectories and orbital separation of the stars, 
respectively.  The gravitational interaction between the two stars has resulted in a decrease in 
the orbital period from its initial value of 1.5 months to less than 5 days.  By comparing the 
time between periastron and subsequent apastron to the time from apastron to subsequent periastron, 
it can be seen that the orbit continues to evolve.  However, the orbital evolution timescale has 
become significantly longer than it was during the early inspiral between 27 and 41 days. In fact, 
the timescale of orbital decay has increased from $\sim 5$ days to $\sim 35$ days. The orbit is also 
gradually becoming more circular with time; its eccentricity at 56.7 days is $\sim 0.08$. We note that 
the periastron separation of $\sim 6 \times 10^{11}$~cm is much larger than the fine-mesh zone spacing
$\Delta x \approx 1.86\times 10^{10}$~cm or the stellar core size, indicating that the orbit is tending
to stabilize at a spatially well-resolved separation.

\begin{figure}[ht]
\centerline{\includegraphics*[width=5.5in]{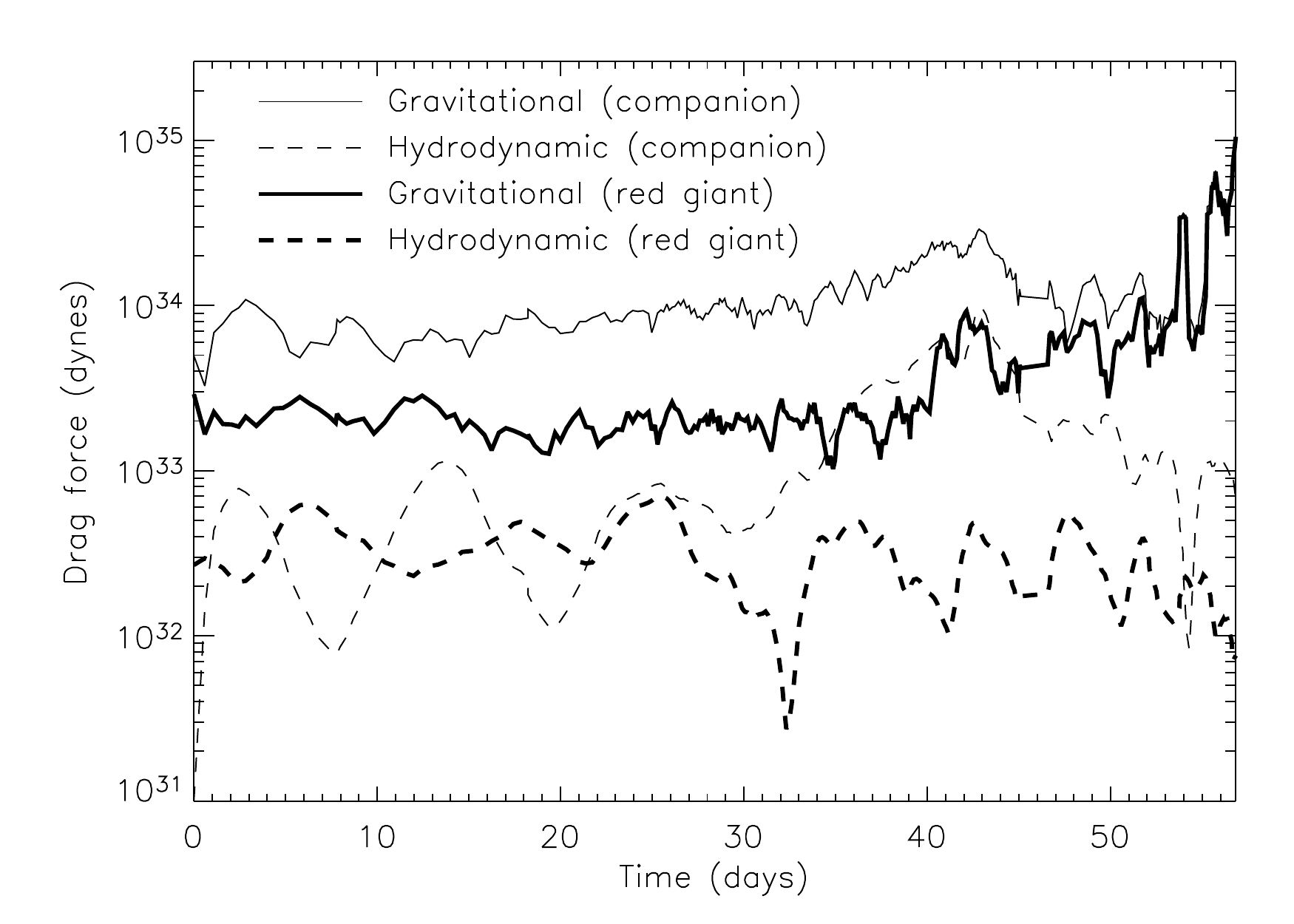}}
\caption{\label{Fig:drag}
         Hydrodynamic and gravitational drag forces acting on the 
           companion star and red giant during the inspiral phase of the CE binary system.
         Force values have been boxcar smoothed over five timesteps.
}
\end{figure}

In order to determine the contribution of the gravitational drag and
hydrodynamic drag acting on the red giant core and the companion star during
the orbital decay phase, we illustrate their variation as a function of time in
Figure~\ref{Fig:drag}.  The hydrodynamic drag force is computed using the
control-surface method described in Paper~I (with a control surface radius of
$3 \times 10^{11}$~cm), and the gravitational drag force is defined 
to be the gravitational force on the particle clouds due to the gaseous envelope.

As for the early stages of CE evolution, the gravitational drag for both
stellar cores continues to exceed the hydrodynamic drag force by more than an
order of magnitude. The ratio of these two forces tends to increase with time
during the late stages of the evolution, primarily because of the increase in
the gravitational drag as the cores spiral together. The gravitational drag on
the red giant core is initially about a factor of 5 less than that on the
companion star because of its central location within the envelope, but as gas
expands to greater radii and the two cores spiral together, the two cores tend
to experience the same gravitational drag force.

\begin{figure}[ht]
\centerline{\includegraphics*[height=3.in]{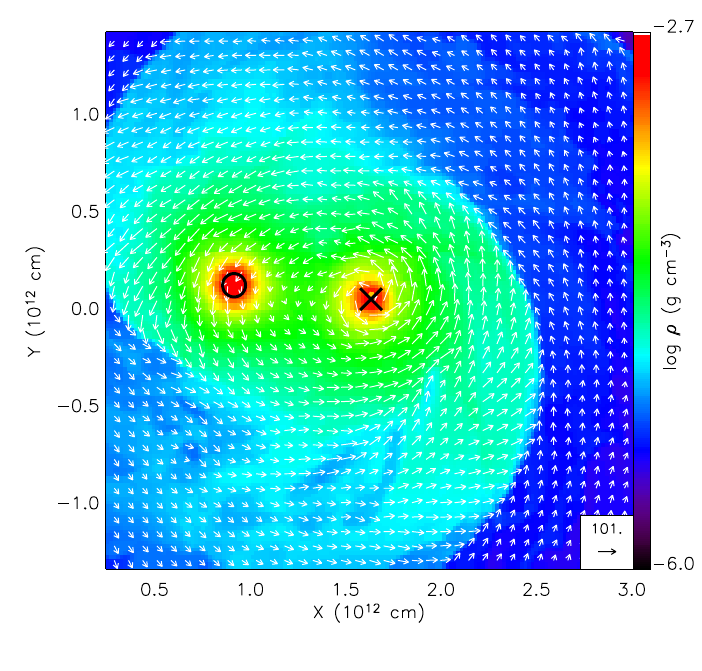}
            \includegraphics*[height=3.in]{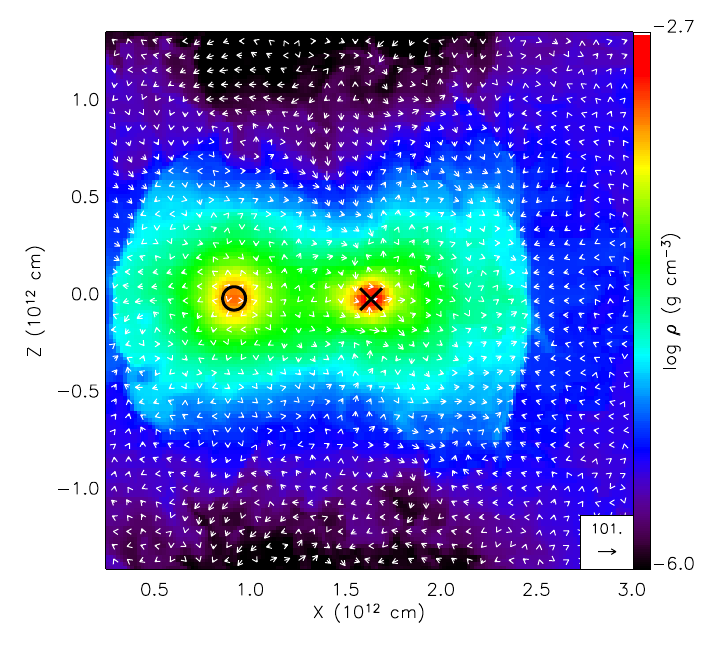}}
\caption{\label{Fig:density}
	 {\it Left:} Density distribution (g~cm$^{-3}$) and velocity field
(km~s$^{-1}$) in the orbital plane at $t = 56.7$~days. The location of the
companion star ($\times$) and red giant core ($\circ$) are shown.
         {\it Right:} As for the plot on the left, but for the $xz$ plane.}
\end{figure}

The hydrodynamic drag force, on the other hand, continues to differ for the two
cores because of the different flow patterns in their vicinities, as can be
seen for the $xy$ and $xz$ planes in Figure~\ref{Fig:density}. The flow in the
vicinity of the red giant core in the orbital plane has been spun up to be nearly uniform; the gas there orbits the
companion with the same velocity as the core. This occurs because of the core's
smaller mass relative to the companion: the center of mass of the system now
lies closer to the companion, and consequently the gas near the red giant core
receives gravitational torques large enough to keep it orbiting with the core.
Because of this approximately uniform flow field, the spherical surface integral of
$\rho {\bold v}_{\rm rel} {\bold v}_{\rm rel}\cdot
{\bold n}$ is small, and little momentum is transferred to the gas
(here $\rho$ is the gas density,
${\bold v}_{\rm rel}$ is the velocity of the gas relative to the red giant core,
and ${\bold n}$ is the normal to the surface).
The companion, on the other hand, is surrounded by a rotating mass of gas, for which
the corresponding surface integral is somewhat larger.

The continued domination of the gravitational drag over the hydrodynamical drag suggests
that the mass accretion rate should continue to be much smaller than the BHL prediction.
We explore this point in the next section.

\subsection{Core accretion}

As discussed in Paper~I, the amount of mass accreted onto the companion and red giant core can be 
estimated by computing the mass flux through a spherical control surface and integrating this flux 
over time. Since we do not attempt to model the accretion process itself, as this would require a 
much higher resolution than we are able to achieve, the radius of the control surface is taken as 
a free parameter.  The choice of this radius should be larger than the spatial resolution to ensure 
convergence and to avoid Cartesian grid effects, but it should also be smaller than the spatial scale 
length of the gas density field to avoid including gas in large-scale flows that are not taking part 
in the accretion process.  Furthermore, the choice in radius is adopted to ensure that the gravity of 
the accretor is the dominant factor affecting the gasdynamics.  The mass flux is determined for a given 
stellar core and control surface radius $R_{\rm acc}$ by choosing $N = 32(R_{\rm acc}/\Delta x)^2$
uniformly distributed points on a sphere of this radius centered on the core's location. The gas density and velocity 
fields (relative to the core's velocity) are linearly interpolated from the AMR grid onto these points, 
and the flux is determined via
\begin{equation}
\label{Eqn:mass accretion}
\dot{M} = \int \rho {\bf v}_{\rm rel} \cdot {\bf n} dA \approx {{2\pi^2 R_{\rm acc}^2} \over N}\sum_{i=1}^N \rho_i {\bf v}_{{\rm rel,}i} \cdot {\bf n}_i \sin \theta_i \ ,
\end{equation}
where the integral is taken over the control surface, $\theta_i$ is the polar angle of the $i$th sample 
point, and ${\bf v}_{\rm rel}$ is the relative velocity of the core and the gas. Note that for the 
control surface radii chosen here the interpolation points all lie within a uniformly refined region of 
the grid.

\begin{figure}[ht]
\centerline{\includegraphics*[height=2.5in]{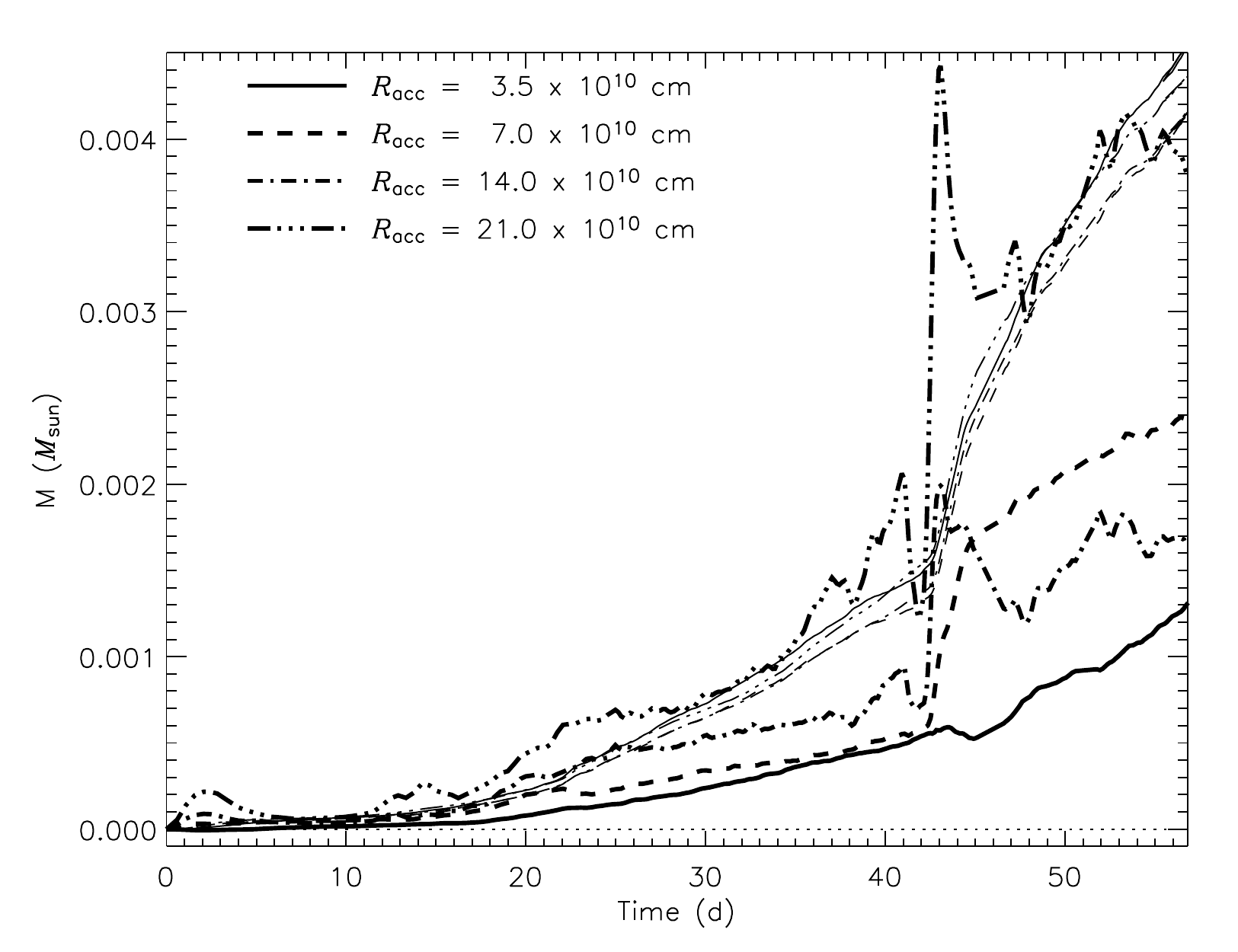}
            \includegraphics*[height=2.5in]{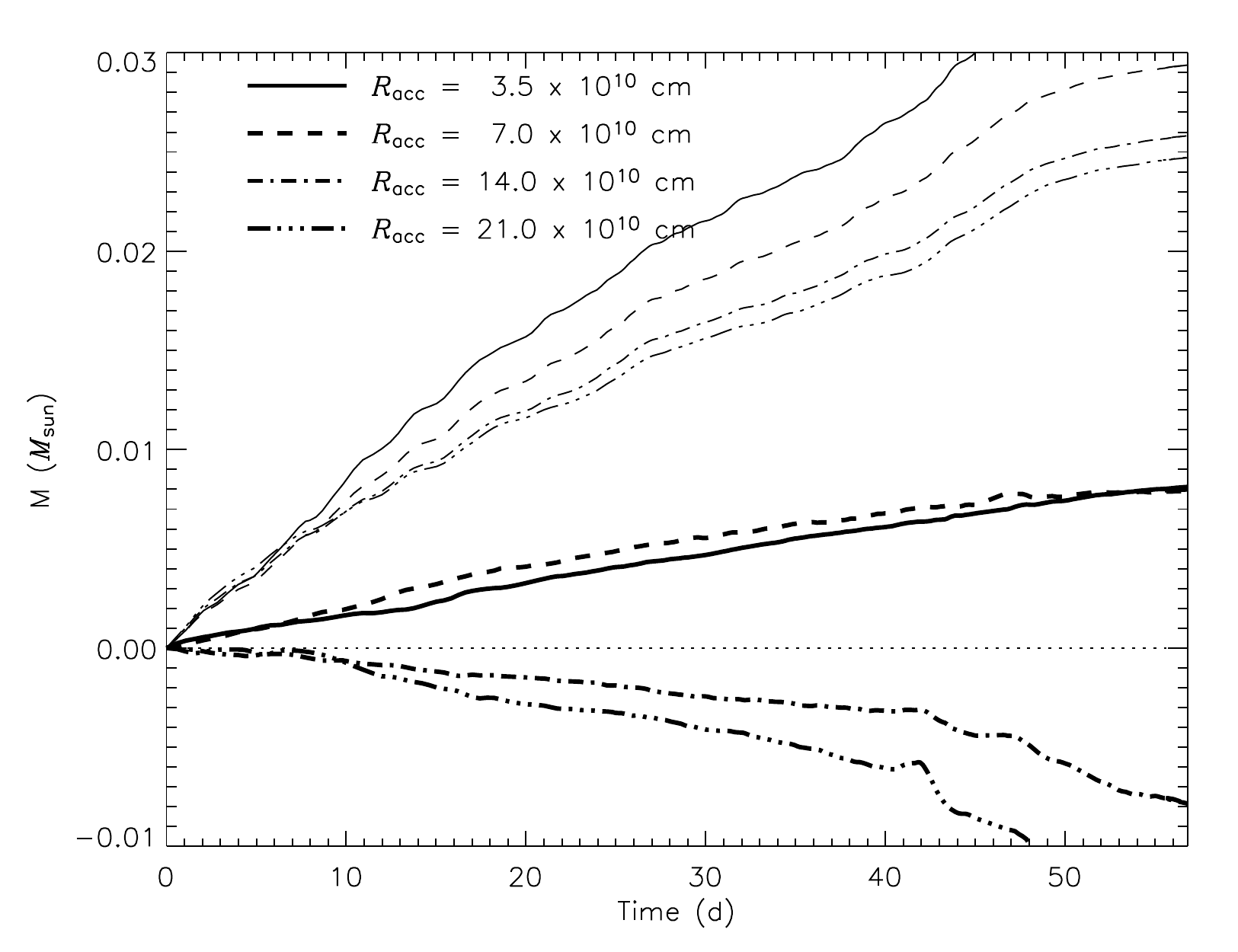}}
\caption{\label{Fig:mass accretion}
{\it Left:} Thick curves indicate mass accretion onto the companion star as determined for control 
surfaces of different radii $R_{\rm acc}$ as a function of time during the inspiral phase of the 
CE binary system. Thin curves show the accreted mass as determined using the BHL model
from density, pressure, and velocity averages within the same control surfaces, divided by 100.
{\it Right:} Same as the left panel, except for the red giant core.
}
\end{figure}

We have performed this calculation for both stellar cores using control surface radii
 between $3.5\times 
10^{10}$ and $2.1\times10^{11}$~cm (1.75 -- 10.5 zones), and the results appear in Figure~\ref{Fig:mass accretion} displayed as a variation of accreted mass as a function of evolution time. For comparison we 
also show the mass accretion to be expected if the accretion rate were determined by the BHL prescription.
In general, the BHL prediction is $\sim 100$ times the amount of mass accreted, as was found for the 
early-inspiral phase in Paper~I.

Our results suggest that the larger choices for control surface radius (1.4 and $2.1\times10^{11}$~cm) 
overestimate the accretion radius as they likely incorporate too much of the cores' 
surroundings to provide a reliable mass accretion estimate. In addition to the disagreement with the 
results for the smaller control surface radii, these choices yield time intervals in which the accreted 
mass decreases, suggesting that the gravity of the individual cores is not the dominant determinant of the 
gas dynamics in regions of these extents.  
The fact that the accreted mass decreases for the red giant core but not the companion is consistent with the fact that the companion's mass is greater than that of the red giant core, so that its region of influence is larger.
The smaller choices (3.5 and $7\times10^{10}$~cm) provide 
better agreement, though after the initial inspiral up to 42~days, they differ by a factor of two for 
the companion.  For the companion, the average mass accretion rate taken from the beginning of the 
simulation is $\sim 10^{-2}\ M_\odot$~yr$^{-1}$, while the higher gas density in the vicinity of
the red giant core yields a larger estimate, $\sim 6\times10^{-2}\ M_\odot$~yr$^{-1}$.

\begin{figure}[ht]
\centerline{\includegraphics*[height=3.in]{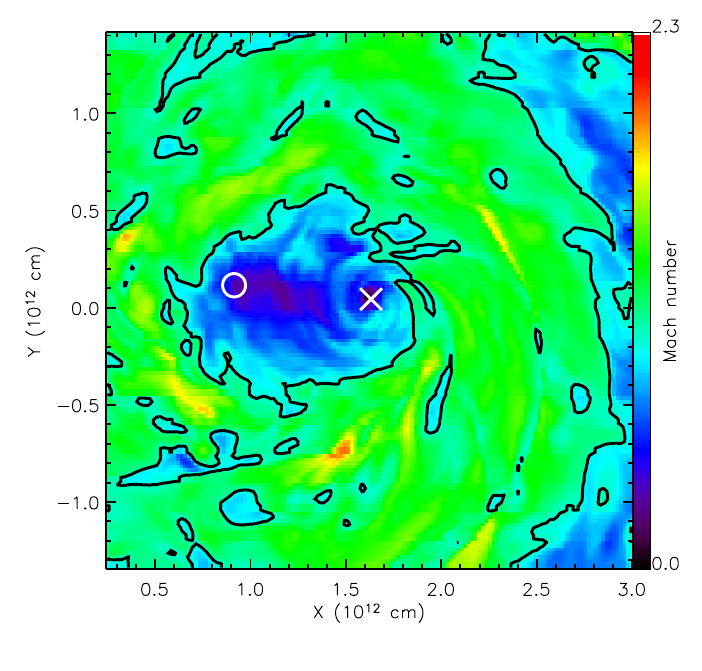}
            \includegraphics*[height=3.in]{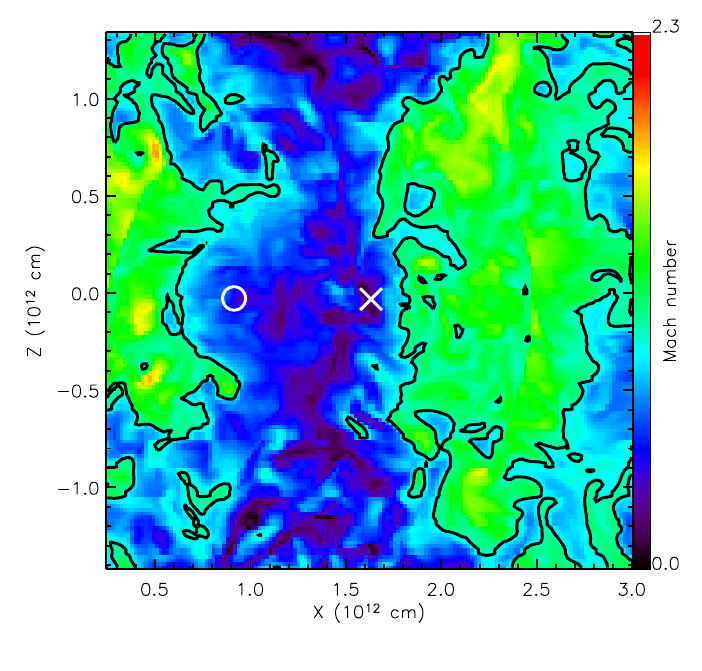}}
\caption{\label{Fig:mach}
	 {\it Left:} Mach number distribution in the orbital plane at $t = 56.7$~days.
The location of the
companion star ($\times$) and red giant core ($\circ$) are shown.
The contour indicates the transition from subsonic to supersonic flow.
         {\it Right:} As for the plot on the left, but for the $xz$ plane.}
\end{figure}

The BHL accretion rate is derived under the assumption that the gas flow is uniform, supersonic, and 
directed against the accretor's direction of travel. We can understand the differences between our 
results and the BHL prediction by examining the flow field in the orbital plane around the red giant core and the companion 
(Figures~\ref{Fig:density} and \ref{Fig:mach}). At late times the flow field continues to differ from 
that assumed in the BHL prescription: not only is the flow subsonic in the vicinity of both cores, but 
for each the flow field is such as to yield a low accumulation rate (uniform across the red giant core, circular about the companion). Thus the assumptions underlying the BHL 
prediction do not hold during the evolution of this system.

Note that since the gas
density is not uniform within the accretion control surface (it varies by an order of
magnitude or so), it might be argued that a BHL rate computed using density, pressure, and velocity values very close to the cores is unrepresentative of the conditions in their
vicinity. However, our estimated BHL accretion rates are based on averages within the
same control surfaces as used for estimating the actual accretion rates. As
Figure~\ref{Fig:mass accretion} shows, the variation in BHL rate
encountered in using different averaging radii is much smaller than the difference
between the BHL rates and the values determined using Equation~\ref{Eqn:mass accretion}.
(Note that the BHL rate curves have been divided by 100 to fit them within the figure.)

\subsection{Outflow}

The angular momentum transported from the orbital motion to the CE via spiral shocks
creates an anisotropic outflow.  Regardless of the outcome of the CE phase (merger or
stabilization as a close binary),
after the complete 
ejection of the envelope the core should contract to the pre-white dwarf stage, evolving to higher 
effective temperatures in the process. If the core is sufficiently massive, the effective temperatures
will be sufficiently high to lead to the creation of a planetary nebula.  In a seminal study, \cite{1987AJ.....94..671B}
suggested that the interaction of a fast stellar wind from the remnant core with an equatorially
dominant circumstellar
envelope may give 
rise to asymmetric planetary nebulae with an elliptical or ``butterfly'' morphology. Although the 
specific model that we have adopted here contains a low-mass helium degenerate core and 
will not produce a planetary nebula, the morphology of the outflowing matter from our simulation is 
generic to outflows resulting from the CE phase \citep[see][]{2000ARA&A..38..113T}.  To provide a quantitative
description of the outflow for potential application to the interacting winds hypothesis, it is 
important to characterize the radial and angular density distribution and the kinematics of the outflow 
resulting from the CE phase. 

\begin{figure}[ht]
\centerline{\includegraphics*[height=3.in]{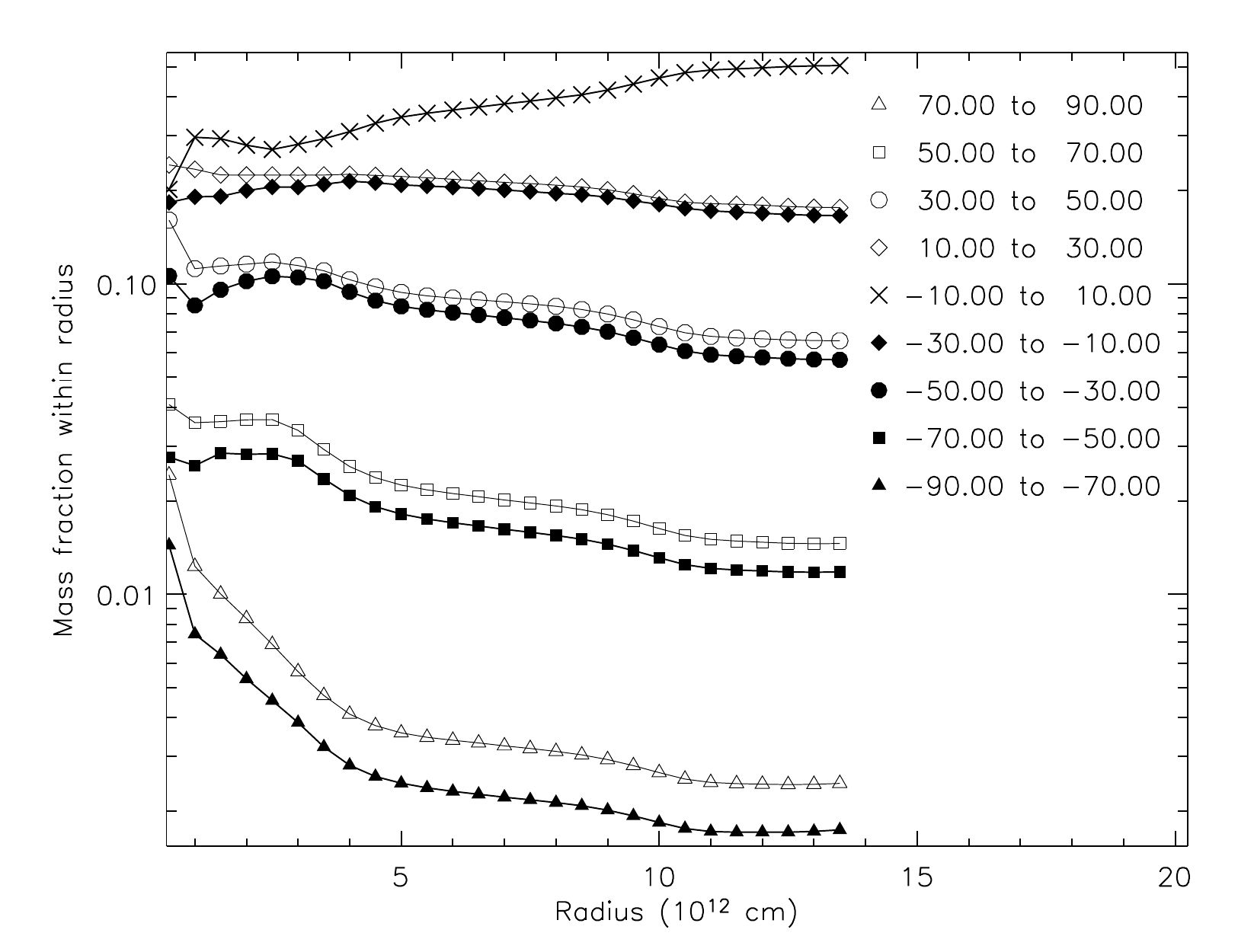}}
\caption{\label{Fig:admoments}
Enclosed mass versus radius for gas lying in different polar angle bins,
as measured using the point $(1.5\times 10^{12}{\rm\ cm},0,0)$ as origin and the Cartesian $z$-axis
as the polar $z$-axis, at $t = 56.7$~d.
}
\end{figure}

Figure~\ref{Fig:admoments} shows the enclosed gas mass as a function of radius about the center of the 
grid for different ranges of polar angle at $t = 56.7$~days. It can be seen that the outflow density 
distribution is equatorially dominant at all radii; however, at small radii (below $10^{12}$~cm) the 
flow is roughly spheroidal. Outside of this region almost 90\% of the outflowing material is confined to an angle of 30$^\circ$ on either side of the equatorial plane.

\begin{figure}[ht]
\centerline{\includegraphics*[height=3.in]{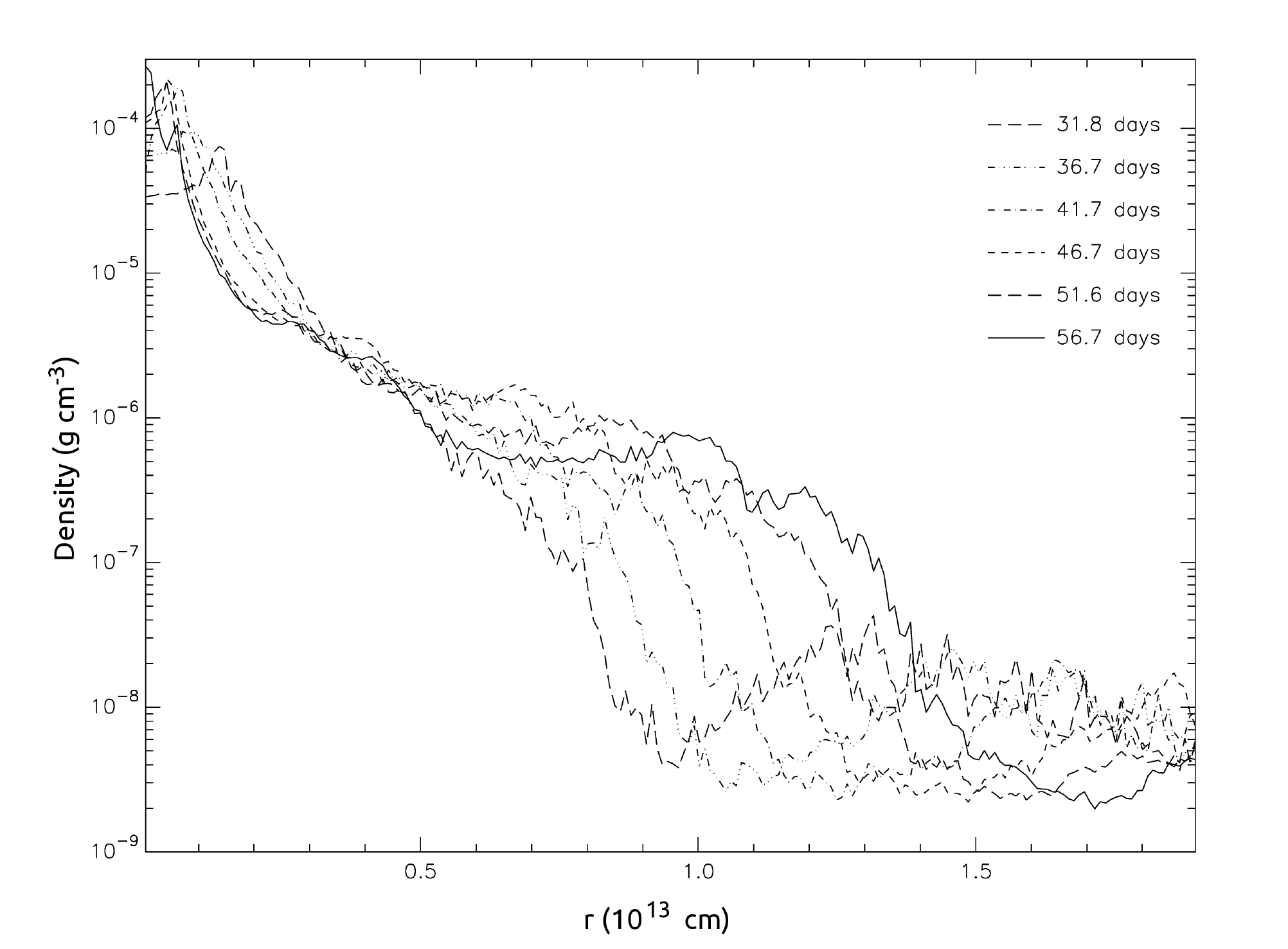}}
\caption{\label{Fig:density-in-plane}
Angle-averaged gas density profile in the orbital plane, using the point $(1.5\times 10^{12}{\rm\ cm},0,0)$ as origin and the Cartesian $z$-axis
as the polar $z$-axis, at different times.
}
\end{figure}

\begin{figure}
\centerline{\includegraphics*[height=3in]{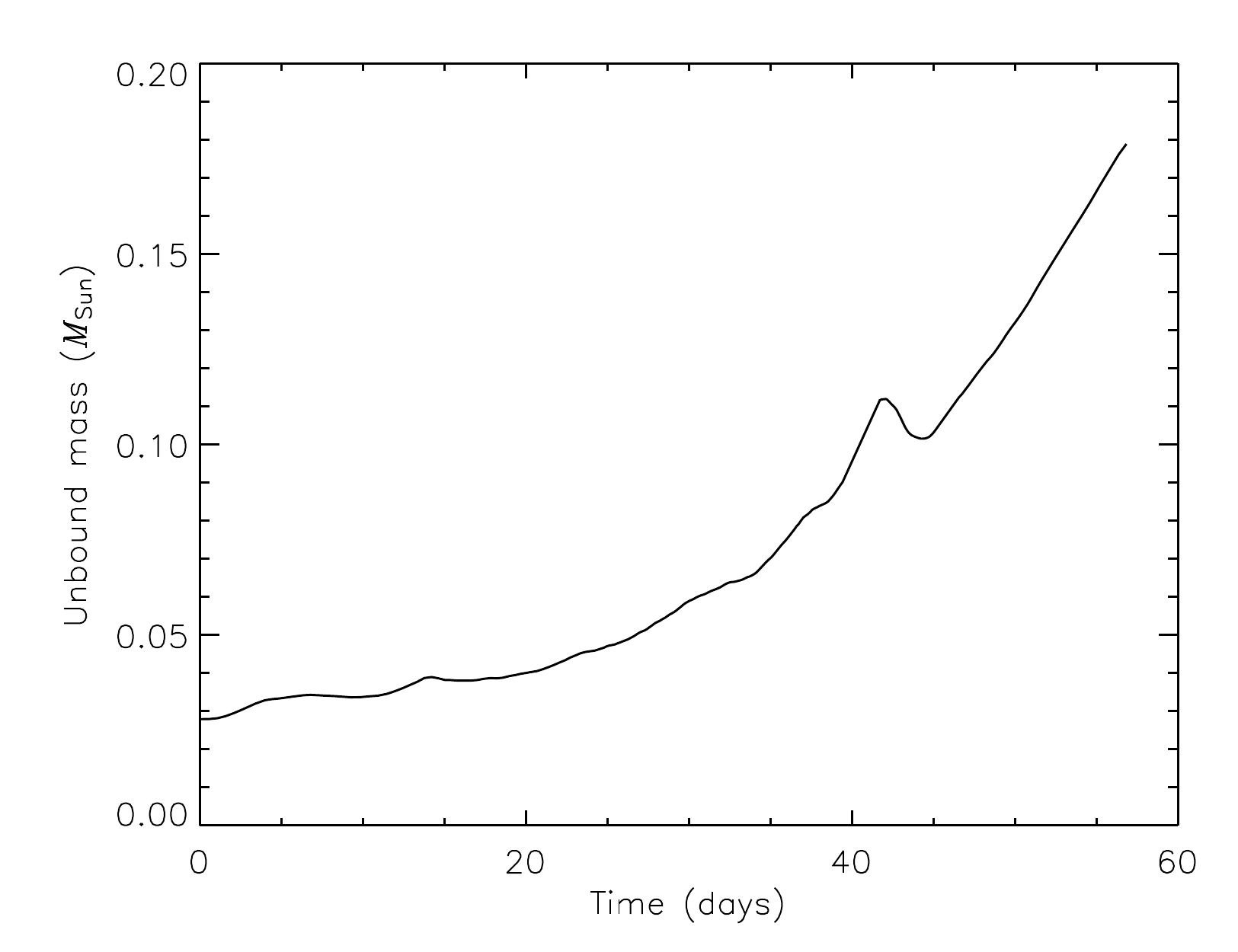}}
\caption{\label{Fig:massloss} Unbound gas mass as a function of time.}
\end{figure}

Figure~\ref{Fig:density-in-plane} illustrates the angle-averaged gas density in the orbital plane as a function of
radius for several different times between 31.8~days and 56.7~days. One can clearly see the expansion 
of the envelope material with time.  Behind the expanding envelope front and outside the spheroidal inner
region the gas density adopts a roughly $r^{-5/3}$ profile. This density slope is intermediate between 
the expectations for constant-velocity isotropic and planar winds, which given the equatorial dominance
suggests that the outflow velocity must increase with radius. The mass loss leads to 
the ejection of $0.18 M_\odot$ of the CE as seen in Figure~\ref{Fig:massloss}, where the amount of 
unbound mass is illustrated as a function of time. The rate of mass loss after about 45 days is nearly 
constant and at the end of the simulation is $\sim 2 M_\odot {\rm \ yr}^{-1}$.

\begin{figure}[ht]
\centerline{\includegraphics*[height=3.in]{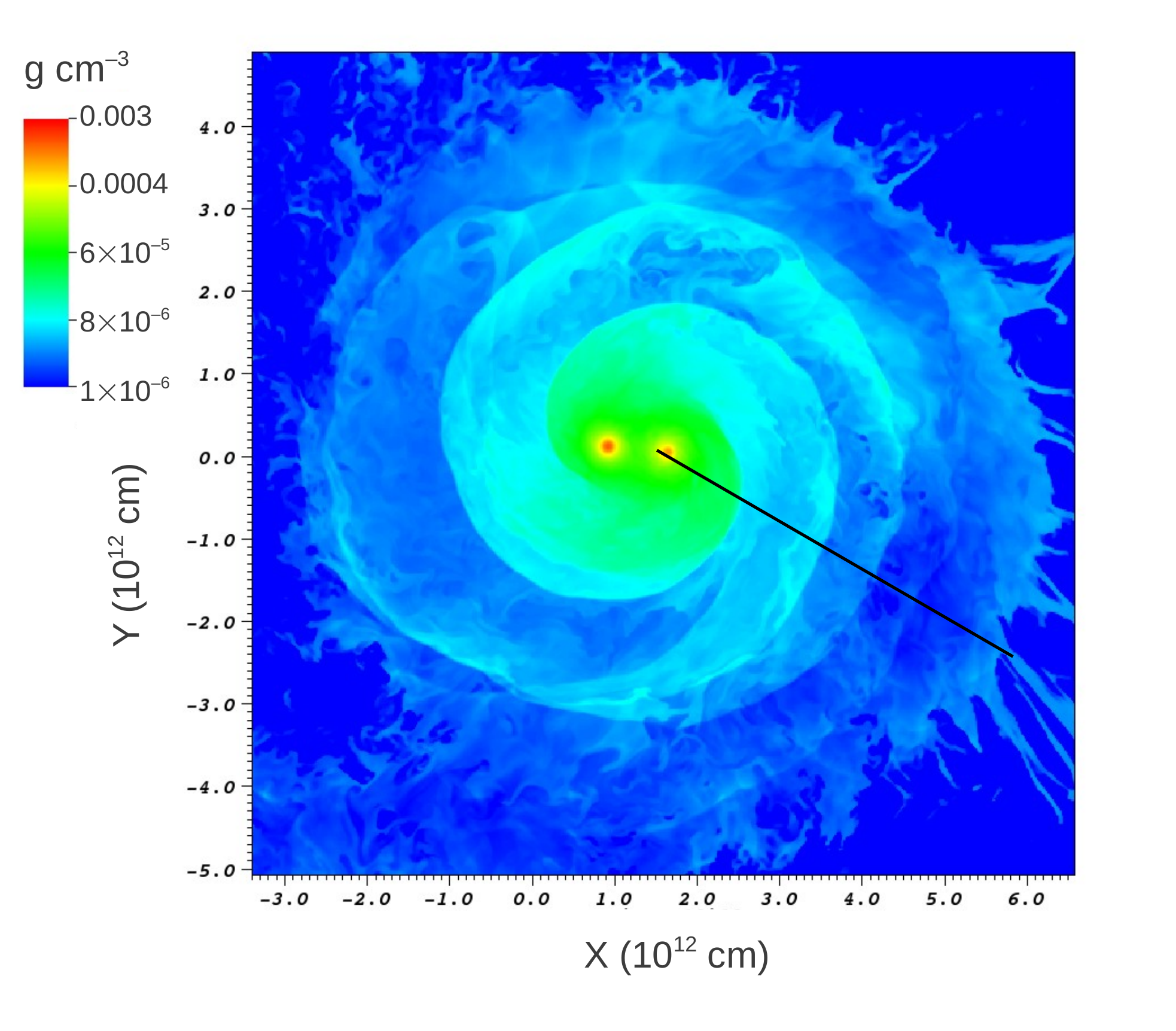}}
\caption{\label{Fig:density-large-scale} Gas density in the orbital plane on large
scales at $t = 56.7$~days. Gas quantities sampled along the black line are plotted in
Figure~\ref{Fig:spiral-shocks}.
}
\end{figure}

\begin{figure}[ht]
\centerline{\includegraphics*[height=4.in]{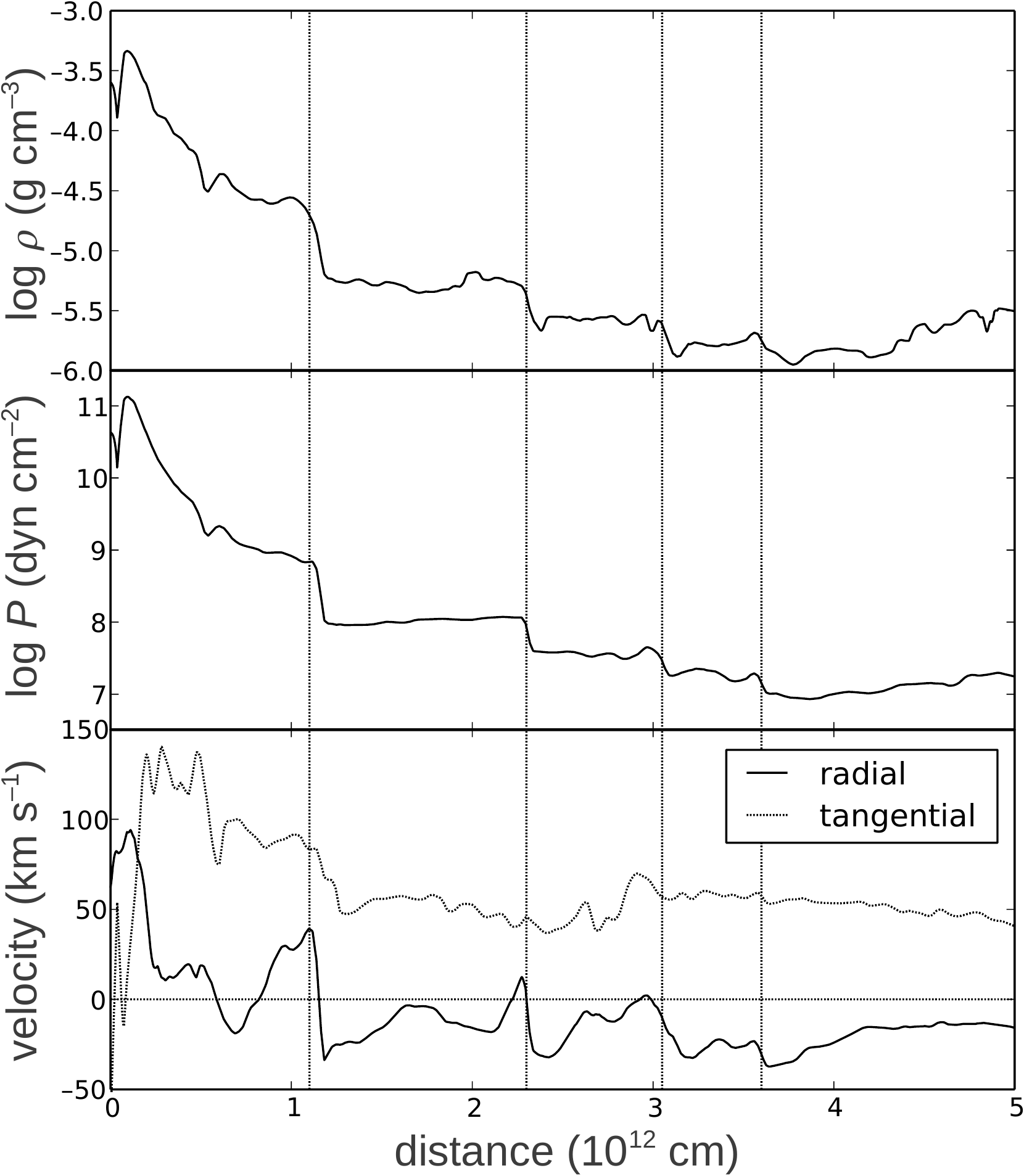}}
\caption{\label{Fig:spiral-shocks} Gas density, pressure, radial velocity, and tangential
velocity profiles along the line indicated in Figure~\ref{Fig:density-large-scale} at $t = 56.7$~days. Vertical dotted lines indicate locations of spiral shocks.
}
\end{figure}

From Figure~\ref{Fig:density}, which displays the distribution of gas in the orbital 
($xy$) plane and $xz$ plane at $t = 56.7$~days, it can be seen that within a radius of about $10^{12}$~cm,
there is no evidence of shock structure, and the gas Mach number is less than one (see Figure~\ref{Fig:mach}).  This suggests that
the angular momentum is removed from the central regions by tidal torques rather than spiral shocks as 
seen in simulations with larger mass ratios \citep[see][]{1998ApJ...500..909S}.  At larger radii, however, 
trailing spiral features are in fact seen (Figure~\ref{Fig:density-large-scale}).  By examining the 
variation of gas quantities along a line segment emanating near the center of mass, we can establish and
quantify the nature of these features. Figure~\ref{Fig:spiral-shocks} shows the variation of gas density, 
pressure, and radial and tangential velocity components along a line in the orbital plane originating at 
the point $(-1.5\times 10^{12}{\rm\ cm}, 0, 0)$ and describing a $-30^\circ$ angle with the $y$-axis. 
This figure clearly shows that the spiral features correspond to jumps in density, pressure, and radial 
velocity, and thus are shocks.  At least four features of decreasing strength can be identified as 
the radius increases. Each feature corresponds to an orbit of the binary following the initial inspiral.

The energetic efficiency of mass outflow, defined as the ratio of the energy
required to unbind the ejected part of the envelope to the energy released by the inspiraling
cores, is of interest in population synthesis models that
incorporate the effects of common envelope evolution. We can estimate the required unbinding
energy by producing radial enclosed mass and energy (kinetic + thermal + potential) profiles for
the initial relaxed system with the companion and red giant spin added
(but before any binary evolution has
occurred). We determine the change in the total bound gas mass from the beginning to the end of
the simulation, giving the amount of mass that is unbound by the stars' interaction.
Assuming that this unbound mass is stripped from the outer layers of the red giant, we use
the mass and energy profiles to find the unbinding energy by subtracting the enclosed energy
at the radius corresponding to the final bound gas mass from the enclosed energy at the
radius at which the binned energy becomes positive.

We find that in the initial conditions only 0.66$\msun$ of the envelope is bound; the remaining
0.03$\msun$ of the envelope is already unbound due to the spin and orbital motion of the red giant.
We include this remaining amount in our estimate of the unbound mass, but not in the calculation
of the efficiency (we can perform the latter by considering only the bound material).
At $t = 56.7$~days, 0.51$\msun$ of the envelope remains bound. The initial
enclosed energies corresponding to enclosed masses of 0.66 and 0.51$\msun$ are $-1.3\times10^{47}$~erg
and $-1.1\times10^{47}$~erg, respectively, so the energy needed to unbind the remaining 0.15$\msun$
that is lost is $\sim 2\times10^{46}$~erg.

At the end of the simulation the 0.51$\msun$ of the envelope that remains bound has an energy
of $-4.6\times10^{46}$~erg, so its energy has actually increased by $6.4\times10^{46}$~erg. However,
because of the inspiral of the stellar cores the total energy of the bound material (gas + cores)
decreases. Using the mesh potential as measured at each core's center, and subtracting the
self-potential of each core, we find that the cores' potential energy decreases by $9.8\times10^{46}$~erg,
roughly consistent with what we expect from the decrease in orbital radius (the difference is due to
the gravitational field of the gas). The cores' kinetic energy
increases by $2.2\times10^{46}$~erg. Thus the total energy released by the inspiraling cores is
$\sim 8\times10^{46}$~erg. Of this amount, 75\% goes into raising the energy of the
part of the envelope that remains bound, and only 25\% goes into unbinding the additional 0.15$\msun$.


\section{Discussion and Conclusions}

We have used a large AMR simulation to study the CE evolution of a binary system initially 
consisting of a 1.05$\msun$ red giant with a 0.36$\msun$ degenerate core and a 0.6$\msun$ 
companion. The use of AMR allows us to simulate stellar pairs with mass ratios close to unity 
at high resolution. We have followed the system through 57 days of evolution, during which 
time it has evolved through five orbital revolutions. The orbital separation decreases by a 
factor of 7 to $8.6 R_\odot$, and the eccentricity steadily decreases with time to a value 
of 0.08. The CE interaction has led to the unbinding of 0.18$\msun$, or 26\%, of the red giant's 
initial gaseous envelope at this point in time.  We note that this fraction of unbound mass is 
larger than in the work by \cite{2000ApJ...533..984S} (typically less than 15\%), but is consistent 
with the trend in which this unbound mass fraction increases with a larger mass ratio (companion 
to the red giant) of the system. 

Based on the release of orbital energy, the efficiency of the mass ejection process is 
$\sim 25\%$. Given that the mass loss rate from the system is $\sim 2 \msun$ yr$^{-1}$, the 
remaining mass is expected to be removed within 2 additional months (comparable to the time 
scale of the orbital decay) provided that the mass loss rate remains approximately constant. This is a reasonable assumption because the binding energy of the portion of the envelope that remains bound is increasing linearly with time at the end of the simulation.
Additional orbital decay will result in the ejection of the remaining envelope.  It is 
possible that ejection of the envelope will be incomplete as some matter may fall back, and 
the effect of any material surrounding the system in the form of a circumbinary disk 
\citep{2011arXiv1105.5698K}
 may lead to further orbital decay.  It is likely that this system will survive 
the CE phase with an orbital period of less than 3 days. 

As in our previous study concerning the early stages of the evolution of this model system, the 
gravitational drag acting on the two stellar cores due to the nonsymmetrical gas distribution 
dominates over the hydrodynamical drag. The flow pattern in the vicinity of the cores in the 
orbital plane at 57~days is nearly uniform for the red giant core and circular for the companion, 
and in both cases it is subsonic. Such a description of the flow renders the assumptions 
underlying simple estimates of the drag based on a BHL prescription suspect and should be noted
when considering evolutionary calculations of lower dimensionality. 

The commonly used BHL prescription for estimating the mass accretion rate onto the companion 
overestimates the actual accretion rate by nearly two orders of magnitude. As discussed in Paper~I, this can have important implications for the likelihood of the possibility 
for accretion-induced collapse of an inspiralling neutron star in a stellar envelope to a 
black hole.  We note, however, that for the less evolved red giant star that we have modelled 
in the present investigation, the rate that we infer from our calculations can still exceed the 
rate for hypercritical accretion \citep[$\sim 10^{-3} \msun$ yr$^{-1}$;][]{1993ApJ...411L..33C} suggesting 
that the outcome of the accretion evolution of an inspiralling neutron star is dependent on 
the evolutionary state of its giant companion. 

An alternative prescription for the CE drag and resulting inspiral has been proposed by Meyer and Meyer-Hofmeister (1979; MM)\nocite{meyer_formation_1979}, who consider a model for the CE evolution of a 5~$M_\odot$ red giant and a 1~$M_\odot$ main-sequence star. In this model, the system is treated as an inner rigidly rotating binary system frictionally interacting with an outer envelope undergoing differential rotation. The angular momentum transfer and energy dissipation in the envelope derives from turbulent convection. Gravitational drag is neglected in the envelope. Because gravitational drag appears to be the dominant dissipation mechanism in our calculation, we would argue that at least during the rapid inspiral phase the MM model greatly underestimates the drag and hence overestimates the inspiral timescale (see their Figure~5). Moreover, the radially averaged specific entropy profile of our system at $t = 56.7$~days is convectively stable, so if convective turbulence should have developed but did not because of our resolution, its integral scale and hence effective mixing length should be smaller than our minimum zone spacing ($2\times10^{10}$~cm). The mixing length in MM is of order the pressure scale height, which in our calculation is $\sim 7\times10^{10}$~cm at its smallest. Hence, the turbulent viscosity is correspondingly overestimated in their model. In our Figure~\ref{Fig:density} it is clear that the velocity field in the central regions is laminar on these scales. Instead, in our calculation angular momentum is removed from the envelope interior regions by spiral shocks.

It is conceivable that physics not included in our simulation could help to invert the entropy gradient and thus drive turbulence on small scales, changing the nature of the angular momentum transport. Our equation of state includes radiation pressure, but the gas is treated as being fully ionized, so energy derived from recombination is not included. If significant amounts of recombination were to occur within the inner regions, it could both raise the entropy there and contribute to the ejection of more envelope material. However, a rough estimate of the optical depth due to electron scattering (which dominates at the temperatures present in the inner regions) shows that gas and radiation remain tightly coupled at least up to a radius of $5\times 10^{12}$~cm. Thus recombination should be important only in the very outer regions of the envelope and not in the inner region where the gravitational binding is greater. In more evolved red giants, where more of the mass is located at larger radii, recombination may play a greater role in unbinding mass, though in such cases one would need to consider the efficiency with which recombination photons couple to the surrounding gas.

The existence of distinct shells of outflowing material corresponding to the different orbits
suggests a means for detecting systems that are undergoing or have recently undergone the common
envelope phase. Once the outflow expands sufficiently to become optically thin, any spectral lines
emitted by the gas will display Doppler-broadened profiles containing discontinuous jumps or features
corresponding to the velocity jumps seen in Figure~\ref{Fig:spiral-shocks}. 
The highest-velocity components result from the tangential motion at radii corresponding to several times 
the orbital separation.  As the outflow expands further, a more radial flow at larger distances is 
expected. Such components would be dependent on the orbital phase of the system and evolve secularly 
with time. These jumps should be
detectable by spectroscopic observations with resolving powers ($\lambda/\Delta\lambda$) of a few thousand
or better, easily within the capabilities of current technology. The primary difficulty with such
observations would be the short time available before the outflow density drops below the detection
threshold. Discovery of post-common-envelope systems using this technique would therefore require
a spectroscopic monitoring program following a large number of stars, or spectroscopic followup of
optical transients detected in a large-area photometric survey with a high cadence (such as LSST).


\acknowledgements{

We acknowledge helpful conversations with Ronald Webbink, Orsola de Marco, and
Jean-Claude Passy, as well as useful comments by the anonymous referee.
Partial support for this work has been provided by NSF
through grants AST-0200876 and AST-0703950.  Computations were carried out
using NSF Teragrid resources at the National Center for Supercomputing
Applications (NCSA) and the Texas Advanced Computing Center (TACC)
under allocations TG-AST040024 and TG-AST040034N.  PMR acknowledges the Kavli
Institute for Theoretical Physics, where some of this work was performed with
funding by NSF under grant PHY05-51164 (the report number for this paper is
NSF-KITP-11-085).  FLASH was developed and is maintained largely by the
DOE-supported Flash Center for Computational Science at the University of
Chicago.

}


\bibliography{cebib}

\end{document}